# ENERGY-BASED SEISMIC DESIGN: NEEDS OF ENERGY DAMAGE INDEX VALUES FOR SERVICEABILITY AND ULTIMATE LIMIT STATES FOR GRAVITY DESIGN BUILDINGS?

**Caterina Negulescu**[1] **and Kushan K. Wijesundara**[2]

[1] BRGM (French Geological Survey), 3 Avenue Claude Guillemin, 45060 Orleans Cedex 2, France
e-mail: c.negulescu@brgm.fr

[2] Faculty of Engineering, University of Peradeniya, Peradeniya, Sri Lanka
kushan@civil.pdn.ac.lk

**Abstract**

*During the past earthquakes, different low ductile failure modes are observed in the gravity design structures and thus, the most of existing damage indices may fail to assess the damage of gravity design structures accurately in referring to the two main performance levels: immediate occupancy and ultimate limit state. Therefore, this study investigates the energy dissipated by the brittle structures and the possible damage indices based on energy for the damage assessment of gravity design frames. In the framework of an Energy-Based Seismic Design Approach, we need the assessment of the Demand and on the Capacity, both expressed in Energy. A methodology for the assessment of the seismic energy demands imposed on structures is already proposed, but not such methodology that makes consensus is proposed for the calculation of the Energy dissipation Capacity avoiding the Hysteretic models. The estimation of the energy expended by the building during an earthquake excitation is a tricky issue. For this purpose, this study considers the wavelet based energy estimation and compares it with different approaches for measuring the damages of a structure: the dominant inelastic period of a building and the more classical measure, the inter-story drift. IDA analysis are performed in energy, drift and inelastic period. Furthermore, the damage assessment results based on the expended energy for three gravity design buildings are compared and discussed relatively to the results expressed in inelastic period and drift. Finally, this study concludes that no significant effects of number of inelastic cycles to the damage assessment results for low ductile structures. However, this study also highlights the effects of number of inelastic cycles to the damage for medium and high ductile structures.*

# 1 INTRODUCTION

During the past earthquakes, different low ductile failure modes were observed in gravity design concrete frame structures. Joint failures, flexural failures, shear failures and combined failure of shear and flexure of mostly the column elements are common types of failure modes ([1] Saatcioglu et al., 2001). In particular, shear failures are observed in the short columns as shown in Figure 1. Such short columns are formed due to the opening placed to accommodate the windows. Therefore, the challenge is which damage index proposed in the literature is suitable to quantify the damage state of such structures more accurately. In the following text, it is briefly discussed the proposed damage indices in the literature.

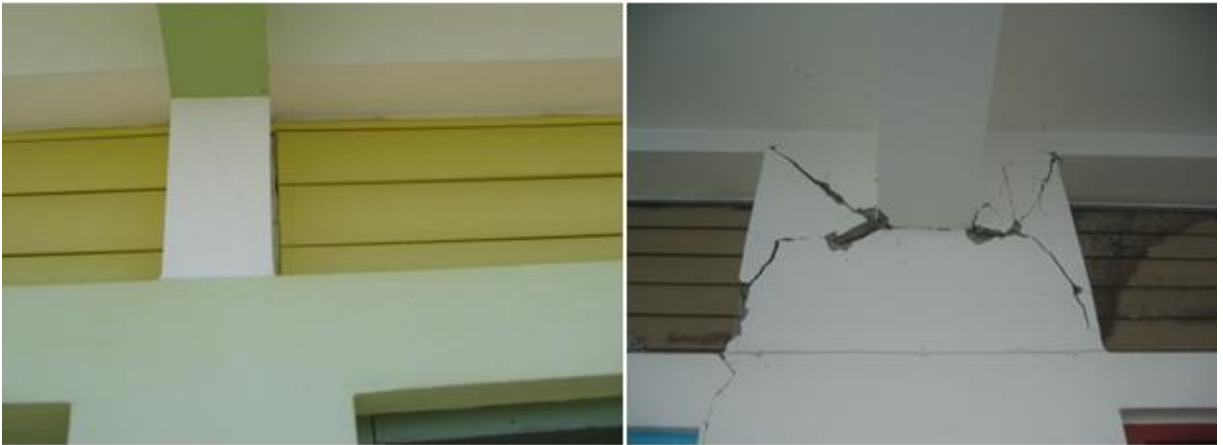

Figure 1: Diagonal cracks in the short columns

A various damage indices have been proposed in the literature to assess the damage state of a structure subjected to a seismic excitation. The studies by [2] Cosenza et al. (1993) and [3] Bozorgnia and Bertero (2001) have summarized the many of the damage indices proposed in the literature. Basically, all the damage indices can be categorized into two different groups depending upon the damage index parameter or parameters used to define the index. They are called non-modal and modal parameter based damage indices. Furthermore, the common feature of many of those damage indices is that they are equal to zero when a structure remains in the elastic range during a seismic event while they are equal to 1 at the complete collapse of a structure.

Non-modal parameters based damage indices are defined by either using a damage parameter such as ductility, which can be defined in terms of curvature, rotation or displacement, inter-storey drift and energy or combination of few of those parameters. The ductility and the interstorey drift are the most commonly used non-modal damage parameters. [4] Powell and Allahabadi (1988) proposed a damage index based on the ductility defined in terms of displacement as expressed in Eq1.

$$DI = \left(\frac{u_{\max} - u_y}{u_{mon} - u_y}\right) = \left(\frac{\mu_{\max} - 1}{\mu_{mon} - 1}\right) \qquad (1)$$

where the $u$ and $u_y$ are the maximum and yield displacement, respectively.

$\mu_{max}=u_{max}/u_y$ is the displacement ductility imposed by an earthquake and $\mu_{mon}=u_{mon}/u_y$ is the monotonic ductility capacity of the structure.

When the ductility is defined in terms of the top displacement of a multi degree of freedom frame, this damage index fails to identify the concentration of damage in a single story.

Therefore, the inter-story drift damage index is used as a better non-modal parameter based damage index to quantify the damage of a structure. It is defined as the ratio of maximum inter-story drift at the center of mass to the ultimate inter-story drift, which usually corresponds to the 30% strength drop of the whole story, as given in Eq 2.

$$DI = \left(\frac{ID_m}{ID_u}\right) \quad (2)$$

Since the ductility and the drift based damage parameters do not account themself the accumulation of damage due to the number of inelastic cycles that the structure is subjected and the energy dissipation demand, they could not estimate the actual damage state of a structure ([5] Mahin and Bertero (1981); [6] Mahin and Lin (1983)). However, it should be noted that drift based damage index could yield good results when assessing the damage state of a structure subjected to a near field seismic event which usually produces a single plus of loading causing the large plastic deformation in structural members or short duration events causing less amount of plastic deformations in the members but with few number of inelastic cycles. Moreover, they are the most commonly used damage indices by engineers and researches due to their simplicity in the estimation of the global damage state of a structure.

Another widely used damage index is the [7, 8] Park and Ang (1985, 1987) damage index which is the linear combination of the ductility defined in terms of displacement and the hysteretic energy dissipation as expressed in the following form.

$$DI = \left(\frac{\delta - \delta_y}{\delta_u - \delta_y}\right) + \beta \frac{E_h}{F_y(\delta_u - \delta_y)} \quad (3)$$

$\beta$ parameter is calibrated using the experimental data.

Since this damage index considers the hysteretic energy dissipation, it includes the cumulative effect of repeated cycles of inelastic response to the damage. However, experimental determination of the $\beta$ parameter is difficult and the methodology is not well described as well. Later, [9] Kunnath et al. (1992) have modified the Park and Ang damage index basically by referring the moment-curvature response of plastic hinge region instead of the force-deformation response of a structural member. The modified damage index is expressed in Eq 4.

$$DI = \left(\frac{\varphi - \varphi_y}{\varphi_u - \varphi_y}\right) + \beta \frac{E_h}{M_y \varphi_u} \quad (4)$$

Even though, both Park and Ang damage index and the modified damage index by [9] Kunnath et al. are calibrated for the concrete member experimentally, they might not be appropriate for assessing the damage state of only gravity design structures without proper calibration of β parameter for poorly confined reinforced concrete members.

[3] Bozorgnia and Bertero (2001) have introduced two improved damage indices for generic inelastic single degree of freedom (SDOF) system in combining of displacement ductility and the hysteretic ductility $\mu_H$ which is defined by [5] Mahin and Bertero (1981) as the ratio of hysteretic energy $E_h$ to energy capacity $E_{hmon}$ under monotonically increasing lateral deformation. They are given in Eqs 5 and 6.

$$DI_1 = (1-\alpha_1)\left(\frac{\mu-1}{\mu_{mon}-1}\right) + \alpha_1\left(\frac{\mu_H-1}{\mu_{Hmon}-1}\right) \tag{5}$$

$$DI_2 = (1-\alpha_2)\left(\frac{\mu-1}{\mu_{mon}-1}\right) + \alpha_2\left(\frac{\mu_H-1}{\mu_{Hmon}-1}\right)^{1/2} \tag{6}$$

where $\alpha_1$ and $\alpha_2$ are constants.

Modal parameters such as natural periods, mode shapes and modal damping ratios can also be used as damage parameters for seismic damage assessment of civil engineering structures. They are widely used for structural health monitoring of large civil engineering structures. Damage indices proposed in the literature using a modal parameter is referred in this paper as modal parameter based damage indices. However, the authors could find very few modal parameter based damage indices in the literature. [10] Di Pasquale and Cakmak (1990) have proposed a damage index based on the natural period of the vibration at the undamaged and damaged state of the structure. The damage index is expressed in the following form:

$$DI = 1 - \frac{T_e}{T_d} \tag{7}$$

where $T_e$ and $T_d$ are the natural period of undamaged and damaged the structure, respectively.

Out of the damage indices discussed before, this study considers only the drift and the natural period based damage indices (as given in Eqs. 2 and 7) for the seismic damage assessment of the three buildings selected in this study as the most suitable damage indices. In addition, two new damage indices are also considered in this study based on the wavelet based energy and the dominant inelastic period of the buildings. Furthermore, the damage assessment results of the three buildings from the four damage indices are compared and discussed. It is important note that the changes of the wavelet based energy and the dominant inelastic period as the damage is progressed are estimated using the continuous wavelet transform (CWT) with complex Morlet wavelet. The CWT method is discussed briefly in the following section.

This paper is organized in the following form that in Section 2, the continuous wavelet transform method and the new damage indices are introduced. In Section 3, the building description and the numerical modelling of the building are presented. In section 4, the seismic damage assessment results from the four different damage indices are discussed. This paper is briefly concluded in Section 5.

## 2 CONTINUOUS WAVELET TRANSFORM METHOD AND NEW DAMAGE INDICES

This section introduces a brief description of the wavelet transform and the wavelet energy based damage index. However, authors strongly recommend to readers to refer the key references to understand the theoretical background of the method ([11] Chui C.K., 1992).

CWT can be used to decompose of a function $x(t)$ into frequency-time domain as defined in the following form:

$$W_{(a,b)} = \frac{1}{\sqrt{a}} \int_{-\alpha}^{+\alpha} x(t)\psi^*\left(\frac{t-b}{a}\right)dt \tag{8}$$

where $\psi^*(t)$ and $b$ are the complex conjugate of $\psi(t)$ and the parameter localizing the wavelet function in the time domain, respectively and $W(a,b)$ are the CWT coefficients that represent the measure of the similitude between the function $x(t)$ and the wavelet at the time $b$ and the scale $a$.

The complex Morlet wavelet which is commonly used for continuous wave transform as a basic function can be expressed as in Eq. 9 and its Fourier transform can be expressed as in Eq. 10. The band with parameter $F_b$ is selected in order to optimize the time and frequency resolution.

$$\psi(t) = \frac{1}{\sqrt{\pi F_b}} e^{2\pi i f_c t} e^{-\frac{t^2}{F_b}} \qquad (9)$$

$$\widehat{\psi}(af) = \frac{1}{\sqrt{\pi F_b}} e^{(F_b \pi^2 (af - f_c)^2)} \qquad (10)$$

where $f$ and $f_c$ are Fourier frequency and central wavelet frequency.

## 2.1 Damage index based on wavelet energy

Furthermore, using the CWT method, which decomposes the signal $x(t)$ into time-frequency resolution, wavelet energy for each scale $a_i$ can be estimated as ([12] Minh-Nghi and Lardiés, 2006):

$$E_{a_i} = \sum_j \left| W_{(a_i b_j)} \right|^2 \qquad (11)$$

In other words, the E(ai) is the summation of square modulus of wavelet coefficients over the number of translations $j$ for a given value of scale $a_i$. As a consequence, the total wavelet energy can be obtained as given in Eq. 12.

$$E_t = \sum_i E_{a_i} \qquad (12)$$

Therefore, this study proposes a new damage index based on the wavelet energy as expressed in Eq. 13 to take into account the effects of number of inelastic cycles to the damage:

$$DI = \left( \frac{E_t}{E_u} \right) \qquad (13)$$

where $E_t$ is the total wavelet energy associated with the acceleration response at the top storey of a structure during the seismic excitation and it is estimated using Eqs. 11 and 12. It is important to note that the top storey response is selected to take into account the effects of the maximum applied response to the damage index. Eu is the ultimate energy of the structure. In order to estimate the ultimate energy of a structure, this study performed the incremental dynamic analyses for 14 real ground motions and the ultimate state of the structure is defined as 30% drop of the strength at any storey level. Therefore, Eu is estimated averaging of fourteen values of total wavelet energy Et corresponding to the failure of the structure.

## 2.2 Damage index based on dominant inelastic period

The wavelet ridges are formed at an instantaneous period and time when the period of the response at a time is equal to the period of the dilated mother wavelet. Therefore, the periods of the vibration can be evaluated at the wavelet ridges where the CWT coefficients reach their maximum values. Since the CWT method is capable to decompose a non-stationary response into the time-period domain, the changing of the dominant inelastic period of a structure due to different level of damage can also be evaluated. Furthermore, it is important to note that the damage index given in Eq. 7 is inversely proportional to the damage period and subsequently, it results that damage index reaches 1 at the period of infinity. This may leads to a under estimation of the damage state of a structure. As a consequence of that this study proposes a new damage index based on the dominant inelastic period $T_d$ of the structure as given in Eq. 14:

$$DI = \left( \frac{T_d - T_e}{T_u - T_e} \right) \quad (14)$$

Where $T_e$ is the elastic period of the first mode and $T_u$ is the inelastic period of the structure at the ultimate limit state of the structure. This relation gives the damage index is linearly proportional to the dominant inelastic period.

## 3 BUILDING DESCRIPTION AND MODELLING

Three different gravity design buildings are investigated in this study. The first building is a six story reinforced concrete structure including one under-ground story as shown in Figure 2(a). It could be considered to be symmetric in plan and elevation. The floor plan is approximately rectangle with the dimensions of 45m and 14.5m in length and width, respectively. In the transverse direction, the building has two bays with the equal bay width of 7m while in the longitudinal direction, it has 16 bays with the equal bay width of 2.6m. The height of each story is 3.1m. Furthermore, it is worth to note that there are some interior in-fill walls for partitioning. The second building is 3 story reinforced concrete wall building shown in Figure 2(b). Initially, it was a reinforced concrete frame building but later, it has been retrofitted with lightly reinforced concrete walls. It could also be considered to be symmetric in plan and elevation. The building has one bay of 6.7m width in the transverse direction and the 12 bays with equal width of 4.4m in the longitudinal direction. Altogether, there are six reinforced concrete walls with equal cross section of 4.1x0.2m in the longitudinal direction. They are continuous to the roof with the same cross section. In-fill walls for interior partitioning are mainly in transverse direction.

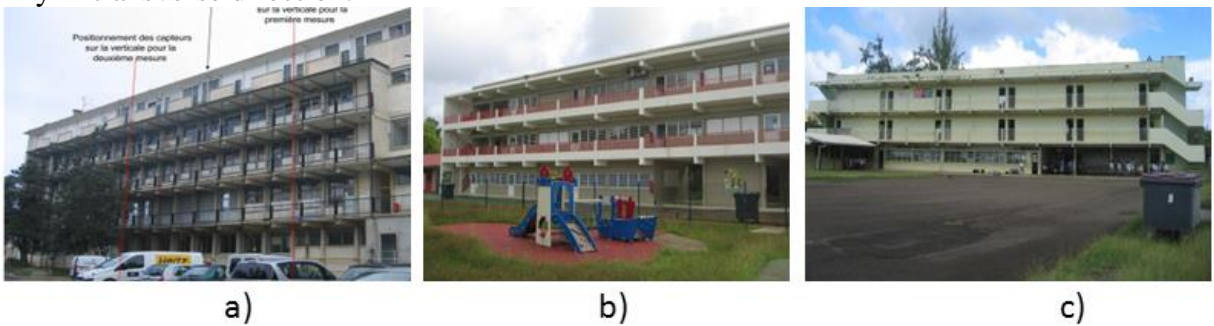

Figure 2: The 3D view of the buildings selected for this study.

Third building is reinforced concrete frame building as shown in Figure 2(c). This building has a single bay of 7.49m width in the transverse direction while it has 10 bays with equal bay width of 4.22m in the longitudinal direction. This could also be considered to be symmetric in the plan and the elevation. It is important to note that there are series of short columns along the longitudinal direction and they are formed due to the presence of openings in the in-fill walls. This building was slightly damaged due an earthquake excitation and all the damages are concentrated in the short columns. The initiation of diagonal cracks in those short columns can be observed as shown in Figure 1(a). However, there is no other damage observed in in-fill walls in either the longitudinal or transverse directions.

A 3-dimensional (3-D) finite element model is developed for each of the building using [13] OpenSees finite element program in order to investigate the performance under seismic loadings. Figure 3(a) illustrates the longitudinal and transversal elevation of the first building. It consists of frame elements and truss elements to represent all the beams and columns, and masonry in-fill walls, respectively. Figure 3(b) illustrates the longitudinal and transversal elevation of the model of the second building which consists of frame elements to represent the beams, columns and the concrete walls. This model consists of the truss elements only in transverse direction. It is important to note that in the first and second building, the masonry in-fill walls are completely filled. Therefore, the truss elements are spanned over the full story height. However, the masonry in-fill walls in the longitudinal direction of the third building are partially filled for accommodating the windows and the doors. Therefore, in the third model shown in Figure 3(c) truss elements are spanned only over the height of the in-fill wall forming the short columns. Furthermore, similar to the first and the second model, it also consists of frame elements to represent the beams and the columns.

All frame (beams and columns) and wall elements in the three models are inelastic beam-column elements available within OpenSees framework. They are based on the force formulation.

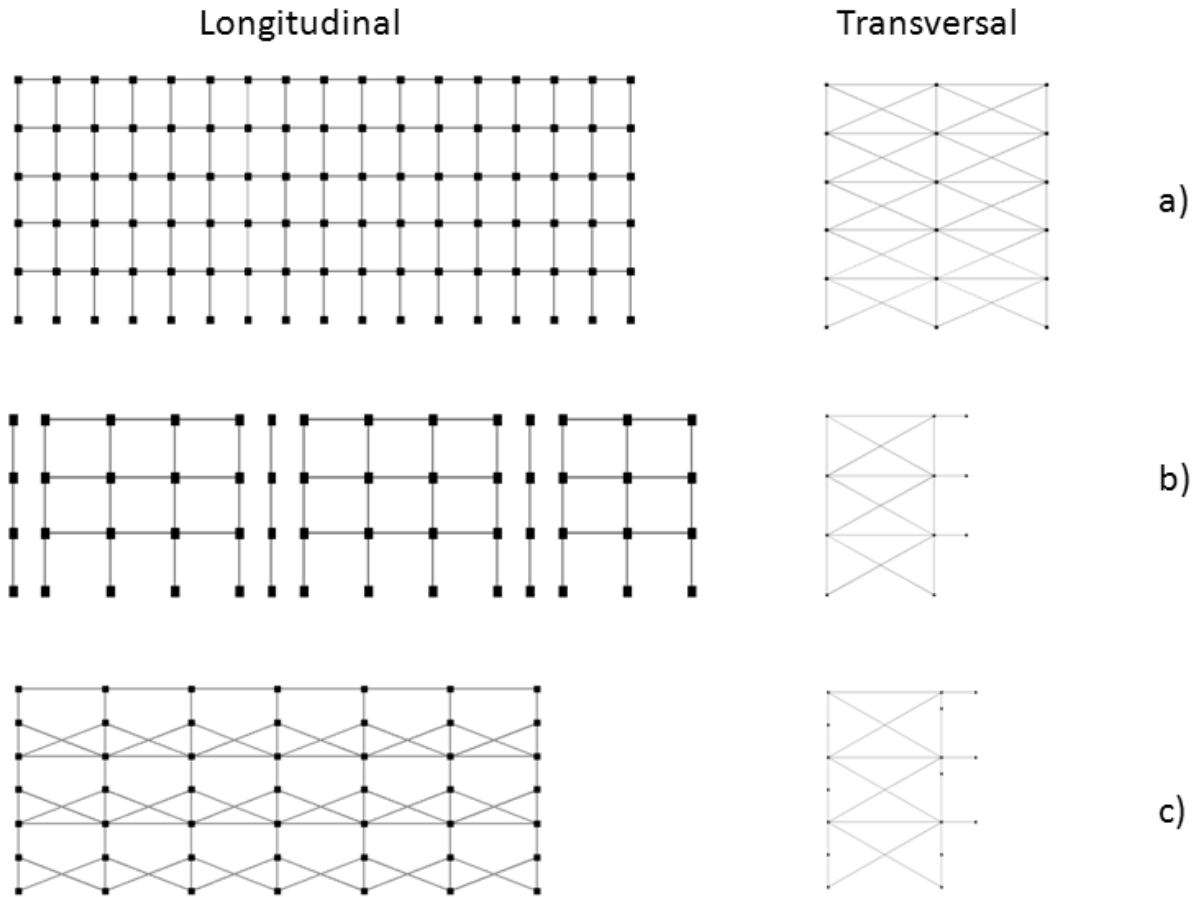

Figure 3. The longitudinal and transversal elevation of the finite element models of the buildings selected in this study.

The material nonlinearity of the concrete represents a uniaxial Kent-Scott-Park concrete material model with degraded linear unloading/reloading stiffness according to the work of Karsan-Jirsa and no tensile strength. Since there are no adequate shear reinforcements provided in the columns and the beams in all three buildings, the confinement effect of the core concrete is minimized.

The masonry in-fill walls in all the three models are conveniently modeled as diagonal struts along its compressed diagonal. The properties of the material and the parameters required to define the geometry of the compressed diagonals are taken as given in [14] Eurocode 8. Figure 4(a) shows the axial force-axial deformation hysteretic response of the truss element during an earthquake. Figure 4(b) indicates the backbone curve used for the concrete short columns with cross section of 0.7x0.2m and the hysteretic response.

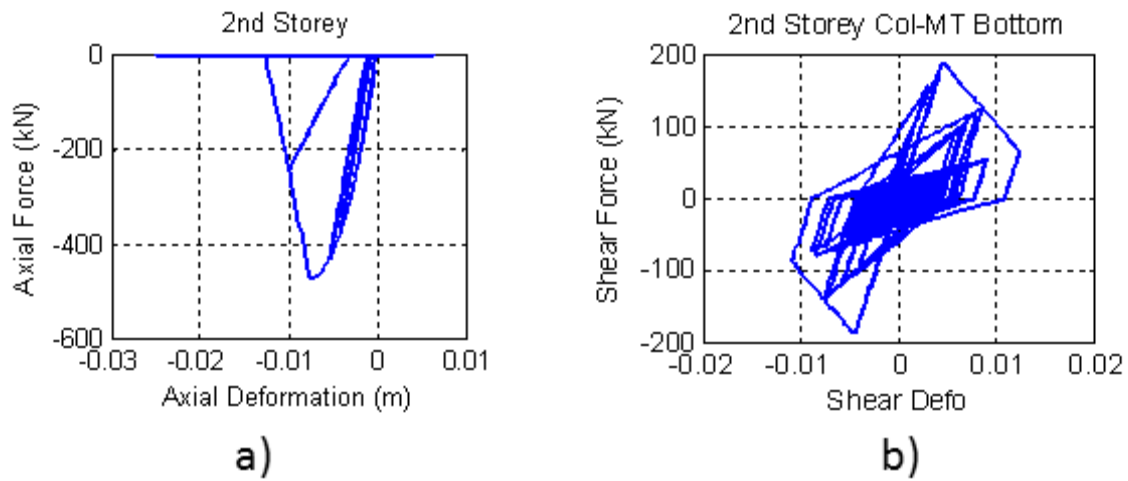

Figure 4. Hysteretic response of (a) axial force-axial deformation (b) shear force-shear deformation.

It is important to note that all the short columns are modeled in taking into account the axial, bending and shear effects. A nonlinear shear force-shear deformation response is attached to the sectional response of the fiber section which accounts the axial-moment interaction. Therefore, the shear deformation is uncoupled from the axial-moment interaction in the section stiffness. However, the shear and bending forces are coupled at the element level because the equilibrium is enforced along the beam element ([15] Marini and Spacone, 2006). The hysteretic response of the shear force-shear deformation is represented by the Pinching4 Material available in OpenSees finite element program. The backbone curve is defined for the concrete section using the modified compression field theory which is implemented in the Response 2000 ([16] Bentz, 2001). Using the Response 2000, the member analysis is performed assuming the fixed support at the bottom of the column while the load on the continuous column. The floor slabs are represented by rigid floor diaphragms ignoring the effect of the flexibility and all the degrees of freedom at the base nodes are assumed to be restrained.

## 4 RESULTS AND DISCUSSION

This study compares the damage assessment results by the four damage indices: the inter-story drift, wavelet energy and inelastic period based damage indices as discussed in the section 1 and 2. As indicated in Eqs. 2, 13 and 14, the three damage indices require the calibrated damage parameters corresponding to the ultimate limit state of the structure. Since they are defined as the ratio of maximum damage parameter imposed by an earthquake to the ultimate damage parameter, the damage indices are equal to 1 at the ultimate limit state of the structure. The first part of this section discusses the adopted procedure for estimating the serviceability and ultimate damage parameters.

### 4.1 Incremental dynamic analysis curves

The promising tool that can be used to calibrate the damage parameters corresponding to the different performance levels is the incremental dynamic analysis (IDA) that has been developed by [17] Vamvatsikos and Cornell (2002). IDA involves nonlinear dynamic analysis of a structural modal under a selected set of ground motions. For this study, 14 real ground motions are selected from the [18] PEER data base. IDA is performed for several scaling levels of each ground motion in order to force the structure to behave all the way from elasticity to its global failure. Subsequently, the IDA curves of structural response are generated as measured by a damage parameter versus the scale factor of the ground motion. The serviceability limit state is defined as an elastic limit of the structure while the ultimate limit state is defined based on the type of the failure mode observed in the critical elements in which larger plastic deformation is expected.

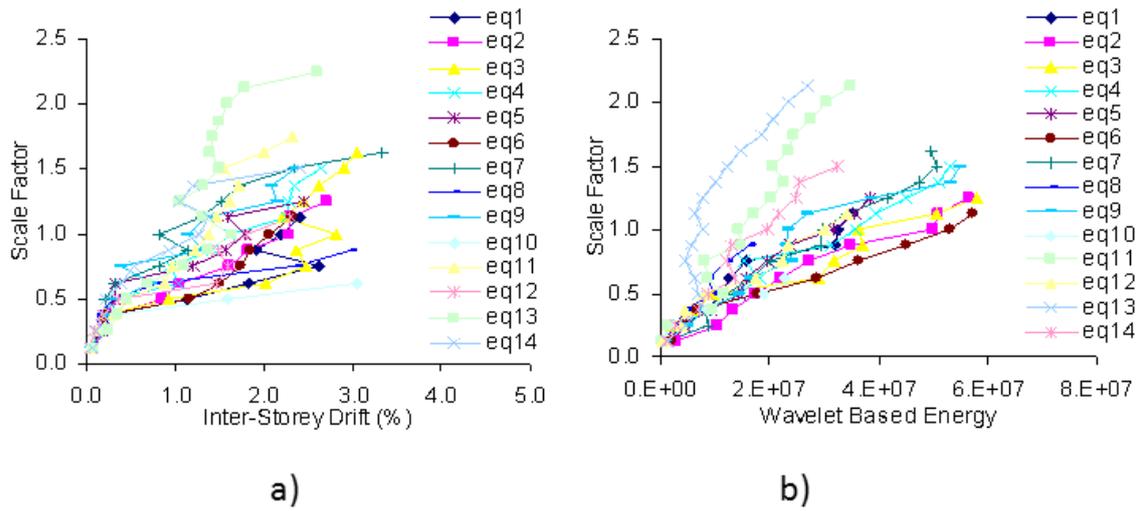

Figure 5. IDA curves for the first building (a) scale factor versus inter-story drift (b) scale factor versus wavelet based energy.

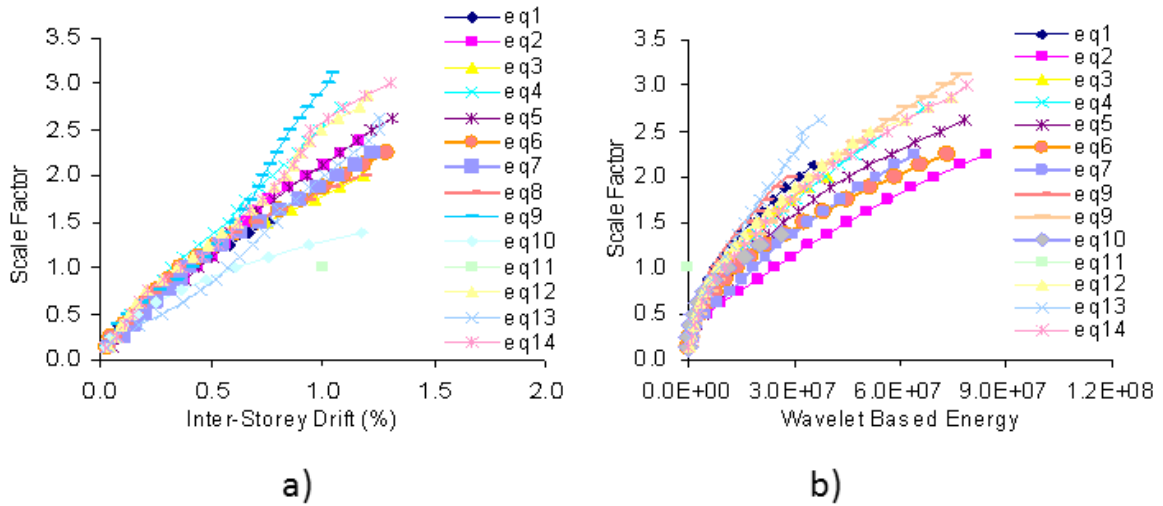

Figure 6. IDA curves for the second building (a) scale factor versus inter-story drift (b) scale factor versus wavelet based energy.

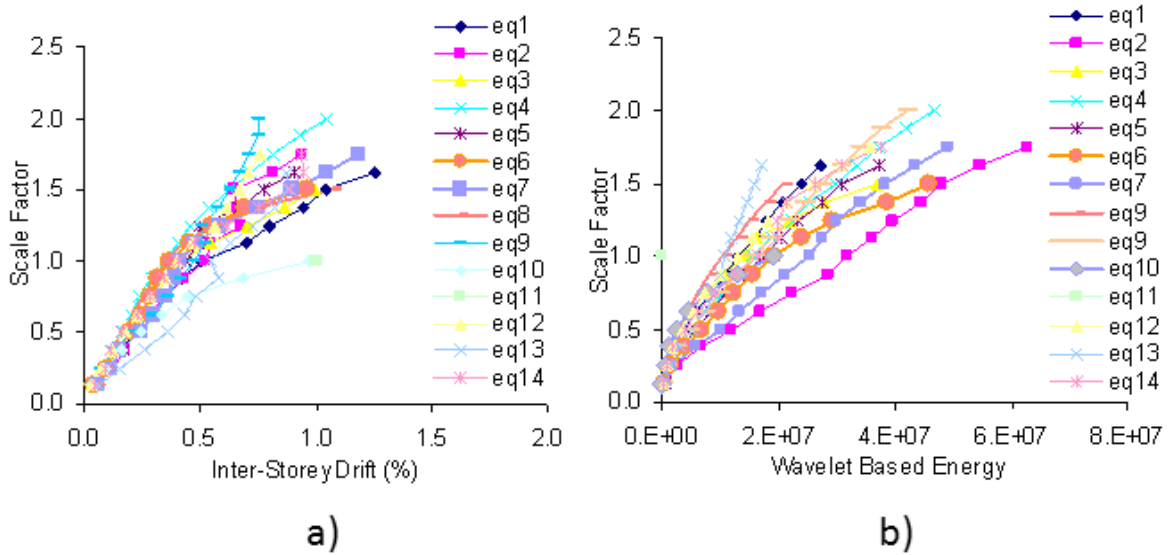

Figure 7. IDA curves for the third building (a) scale factor versus inter-story drift (b) scale factor versus wavelet based energy.

It is clear from Figure 5(a), 6(a) and 7(a) that IDA curves start as straight line in the elastic range and then shows the softening by displaying a tangent slop less than the elastic. Figure 5(a) illustrates significant softening after the initial straight branch of each curve due to compression failure of the diagonal struts which represent in-fill walls at the first storey level. However, some of the IDA curves harden displaying the slope almost equal to the elastic slope even at the higher drift level. Finally, all the IDA curves start softening again showing the larger record-to-record variability due to different characteristics of the ground motions and their effect on the inelastic response. Furthermore, Figure 6(a) and 7(a) also indicate the significant softening displaying the effect of yielding. They also display the record-to-record variability but it is not significant as in Figure 5(a). This could be due to the fact that both the second and the third buildings are subjected to a lower amount of plastic deformations before the failure. IDA curves generated as measured by wavelet based energy versus scale factor shown in Figure 5(b), 6(b) and 7(b) display the gradual increment of the amount of energy dissipation as increasing the scale factor beyond the elastic limit. This proves that wavelet based energy could be a good damage index parameter. Furthermore, it should be noted that the last point of each of the IDA curves corresponds to the failure of the structure. In the next section, it is discussed in details how the failure point on the IDA curve is defined incorporating element failure to the global failure of the structure.

### 4.2 Defining serviceability and ultimate limit state

As discussed before, the serviceability limit state of each structure is defined as the limit of the elastic limit. For the first building, serviceability limit point on each IDA curve is defined corresponding to the failure of the first story in-fill wall while for the second and the third building it is corresponding to the flexural yielding of the first story columns. Table 1 summarizes the average values of the wavelet energy and the inert-story drift based damage parameters corresponding to the serviceability limit state. It is important to note that the elastic period of each of the three buildings are estimated from the ambient vibration measurements using the CWT method as reported in paper by [19] Wijesundara et al. (2012). Furthermore, each of the numerical models is validated by comparing the estimated first mode period from ambient vibration measurements to the first mode period obtained from the Eigen value analysis. Table 1 also gives the measured and the numerical 1st mode periods for the three buildings considered in this study.

| Bld. No | Damage Parameters at Serviceability Limit State | | | |
|---|---|---|---|---|
| | Wavelet energy ($E_t$) | Inter-story Drift, $DR$ (%) | Measured 1st mode elastic Period, T sec | Numerical 1st mode elastic Period, T sec |
| 1 | 11978695 | 0.58 | 0.330 | 0.320 |
| 2 | 10359584 | 0.29 | 0.230 | 0.235 |
| 3 | 17780476 | 0.46 | 0.303 | 0.312 |

Table 1: Damage Parameters at Serviceability limit state.

According to the study by [17] Vamvatsikos and Cornell (2002), the ultimate limit state point on an IDA curve is defined as a point where the IDA slope is equal to 20% of the elastic while it also belongs to a softening branch. However, this study incorporates the element performance to define the ultimate limit state point of the structure on the IDA curve. From the numerical investigation, it is evidenced that the global failure of the first building results in the failure of first story column elements in flexure due to the formation of soft-story mechanism. As the result of the gravity design of the frames, effective depths of the beams are higher than the columns and, in turns, this results beam sections have more strength and stiffness than the corresponding column sections. Therefore, plastic deformations are concentrated at the first story columns forming the soft story mechanism. The failure of the second building results in the failure of wall elements in flexure at their bases. In both the cases, the strength drop of the flexural elements results mainly due to the crushing of core concrete. Therefore, in this study, the failure of individual frame element in flexure is defined by the 30% drop from the peak load at the peak rotation ductility. The cumulative effect of strength drop of the individual elements in any story causes the significant drop of the story shear capacity. The drop of the story shear capacity is approximately equal to the drop of the moment capacity of individual element based on the assumption that all the columns in the story level reach their peak deformation simultaneously. As a consequence of this, the global failure points on IDA curves of the first and the second buildings corresponds to the 30% drop from the peak shear capacity of any story level.

Due to opening placed for the windows and doors as shown in Figure 1, the short columns in the third building has already subjected to an initiation of diagonal cracks leading to the shear failure. An individual element failure in shear is defined as stating of the negative incremental stiffness of the shear force-shear deformation response of the element. Thus, the global failure point of the structure on the IDA curve is defined as all the short columns in the story level reach their individual element failure. Table 2 summarizes the average values for the three damage parameters corresponding to the ultimate limit state.

| Bld. No | Damage Parameters at Ultimate Limit State | | |
|---|---|---|---|
| | Wavelet energy ($E_t$) | Inter-story Drift, $DR$ (%) | 1st mode inelastic Period, $T_u$, sec |
| 1 | 42330660 | 2.68 | 1.13 |
| 2 | 48091036 | 1.20 | 0.789 |
| 3 | 31397630 | 0.97 | 0.702 |

Table 2: Damage Parameters at ultimate limit state.

As specified in [20] FEMA 356 (2000), the inter-story drift limits corresponding to the collapse prevention performance level for well design reinforced concrete frame and concrete wall buildings for seismic loading are 4% and 2%, respectively. The collapse prevention performance level, which is defined as the post-earthquake damage state in which building is on the verge of partial or total collapse, corresponds to the ultimate limit state of the structure. However, the inter-story drift limits in Table 2 is significantly lower than the specified limits in [20] FEMA 356 (2000). This indicates the vulnerability of gravity design structures against a seismic loading.

Table 3 present the damage levels obtained from the wavelet energy and the inter-story drift based damage indices corresponding to the serviceability and ultimate limit states of the three buildings.

| Bld. No | Serviceability Limit State | | Ultimate Limit State | |
|---|---|---|---|---|
| | Wavelet energy damage index | Inter-story damage index | Wavelet energy damage index | Inter-story damage index |
| 1 | 0.28 | 0.22 | 1.00 | 1.00 |
| 2 | 0.22 | 0.24 | 1.00 | 1.00 |
| 3 | 0.57 | 0.47 | 1.00 | 1.00 |

Table 3: Damage levels corresponding to serviceability and ultimate limit states.

It is clear from Table 3 that the damage levels obtained from the wavelet energy and the inter-story drift based damage indices are not significantly different and they are correlated well with the calibration of minor damage state (0.1< Damage Index < 0.25), which is proposed by [7] Park et al. (1985) based on observations of post-earthquake damage of reinforced concrete buildings, for the first and the second buildings for which the ultimate limit state is defined in failure of element in flexure. However, the higher damage level is observed for the third building due to the brittle failure of element in shear. As a consequence, the damage levels for the assessment of a structure using the either of wavelet energy or inter-story drift based damage index must be specified according to the mechanism which leads to the global failure of a structure.

### 4.3 Comparison

In the first part of this section, the correlation between the wavelet energy based damage index and the drift based damage index is investigated based on the results of the IDA. For this purpose, each IDA curve is normalized by its ultimate value of the damage parameter. Figure 8 (a), (b) and (c) illustrate variation of wavelet energy based damage index against the inter-story drift based damage index for the first, second and third building, respectively. The gray solid line in each figure indicates the linear correlation between the damage indices while the black solid line indicates the linear regression curve fitted to the data points.

It is clear from Figure 8(b) and (c) that strong linear correlation between the wavelet energy based damage index and the inter-story drift based damage index can be observed in the second and third structures which have relatively low drift capacity. Therefore, it is proven that the effects of number of inelastic cycles to the damage are not significant for relatively lower deformation levels. However, Figure 8(a) shows highly scattered variation of wavelet energy based damage index against the inter-story drift based damage index due to the fact that the wavelet energy based damage index can account for the effects of number of inelastic cycles and the level of maximum displacement to the damage of a structure. Therefore, this highlights the effects of number of inelastic cycles to the damage for medium level of inelastic deformations. However, the linear correlation between the two damage indices can still be assumed.

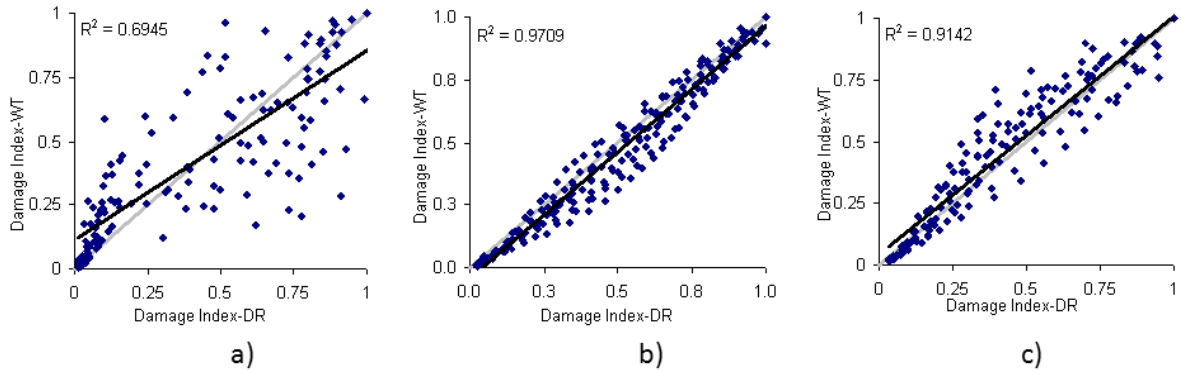

Figure 8. Correlation between wavelet energy based damage index and the inter-story drift based damage index (a) first building (b) second building and (c) third building.

In the second part of this section, the damage assessment results by the four damage indices expressed in Eqs. 2, 7, 13 and 14 are compared. For this comparison, the three structural models are subjected to 50 real ground motions. It must be noted that the ground motion records that have been already used in the previous section are not included in this set of ground motions. Since the structure is symmetric in plan and the elevation, the unidirectional loading in the longitudinal direction (weaker direction) is considered. Figure 9 illustrates the assessment results. Each damage indices used for this comparison is equal to 1 at the ultimate limit state of a structure , therefore the resultant damage index above 1 are considered to be equal to 1.

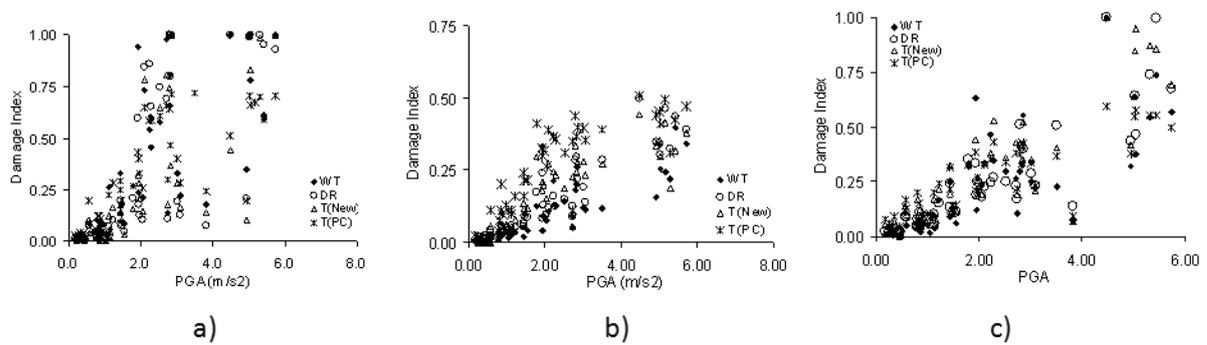

Figure 9. Comparison of damage assessment results by the four damage indices (a) first building (b) second building and (c) third building.

Figure 9 (a), (b) and (c) show variation of damage assessment results by the four equations against the PGA values. However, the main objective of this study is to compare the damage assessment results by the four equations. As discussed before, the damage index based on the inter-storey drift is widely used for the damage assessment of the building by the engineers and the researchers. Therefore, this study further investigates the correlation between the drift based damage index and the other damage indices.

The plots of (a), (b) and (c) in Figure 10, 11 and 12 illustrate the correlation between the inter-story drift based damage index and the wavelet energy based damage index, inelastic period based damage index as proposed in this study and the inelastic period based damage index as proposed by [10] Di Pasquale and Cakmak (1990), respectively while Figure 10(d), 11(d) and 12(d) illustrate the correlation between the period based damage index proposed in this study and the wavelet energy based damage index for the three different buildings considered in this study.

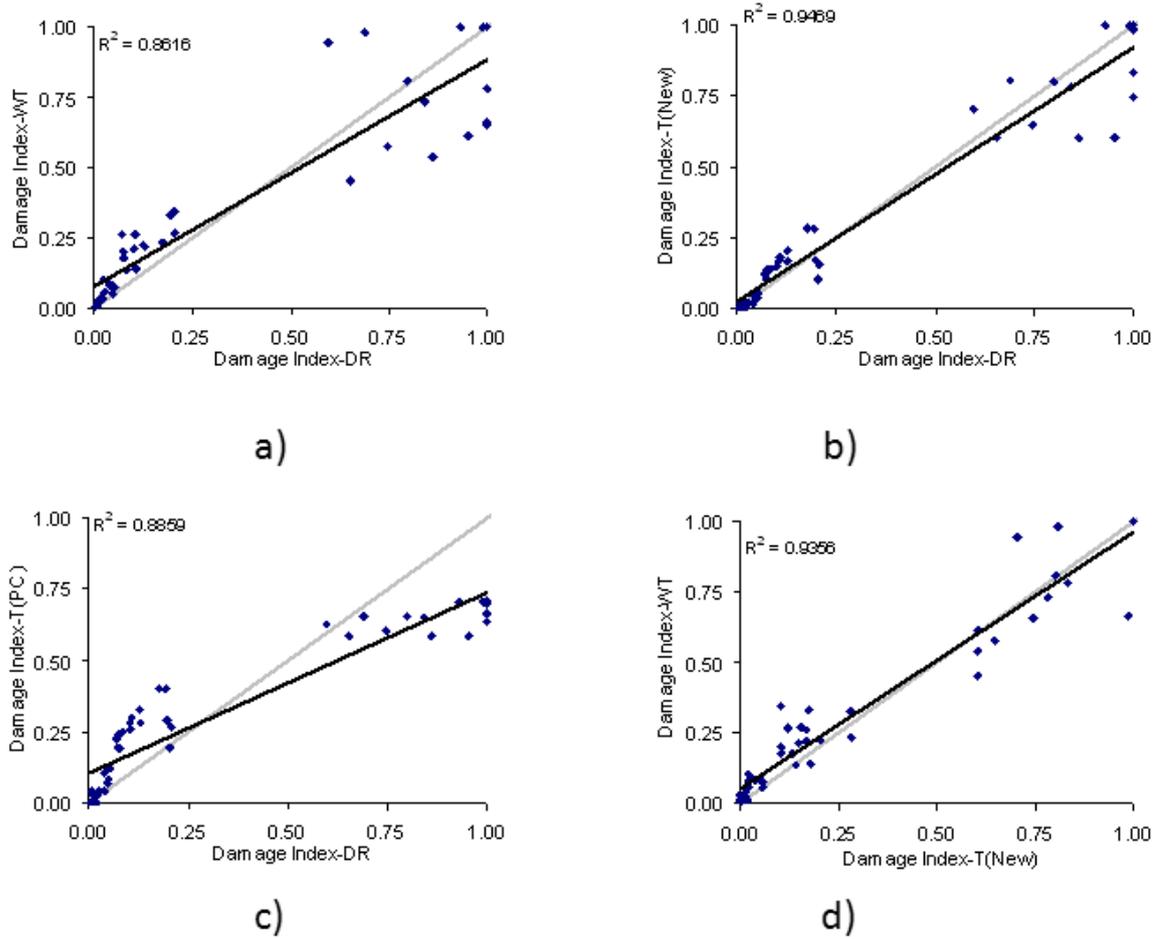

Figure 10. Correlation between different damage indices reference to the first building.

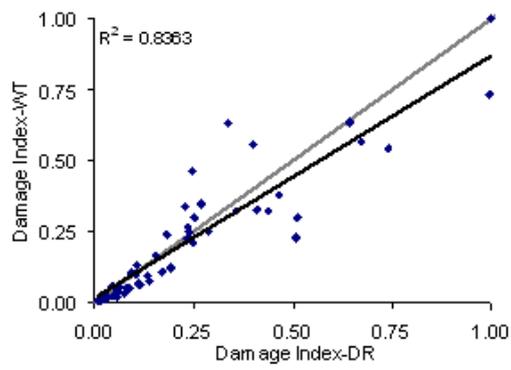

a)

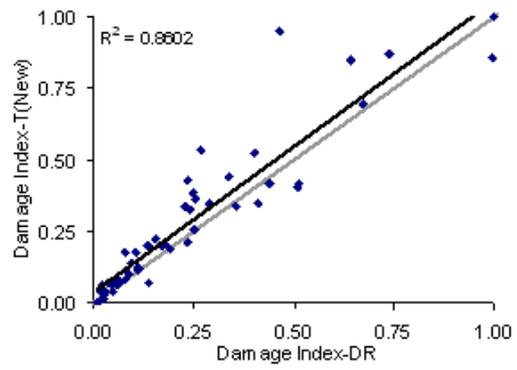

b)

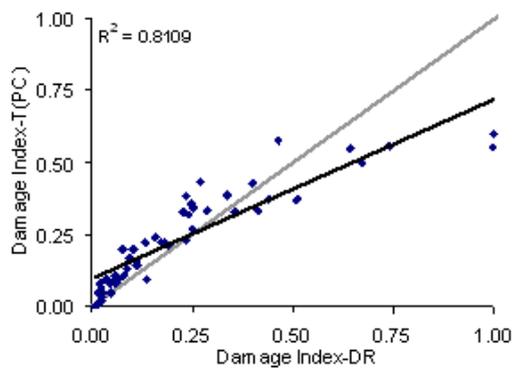

c)

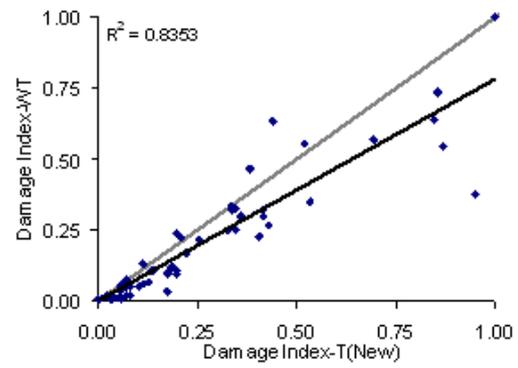

d)

Figure 11. Correlation between different damage indices reference to the second building.

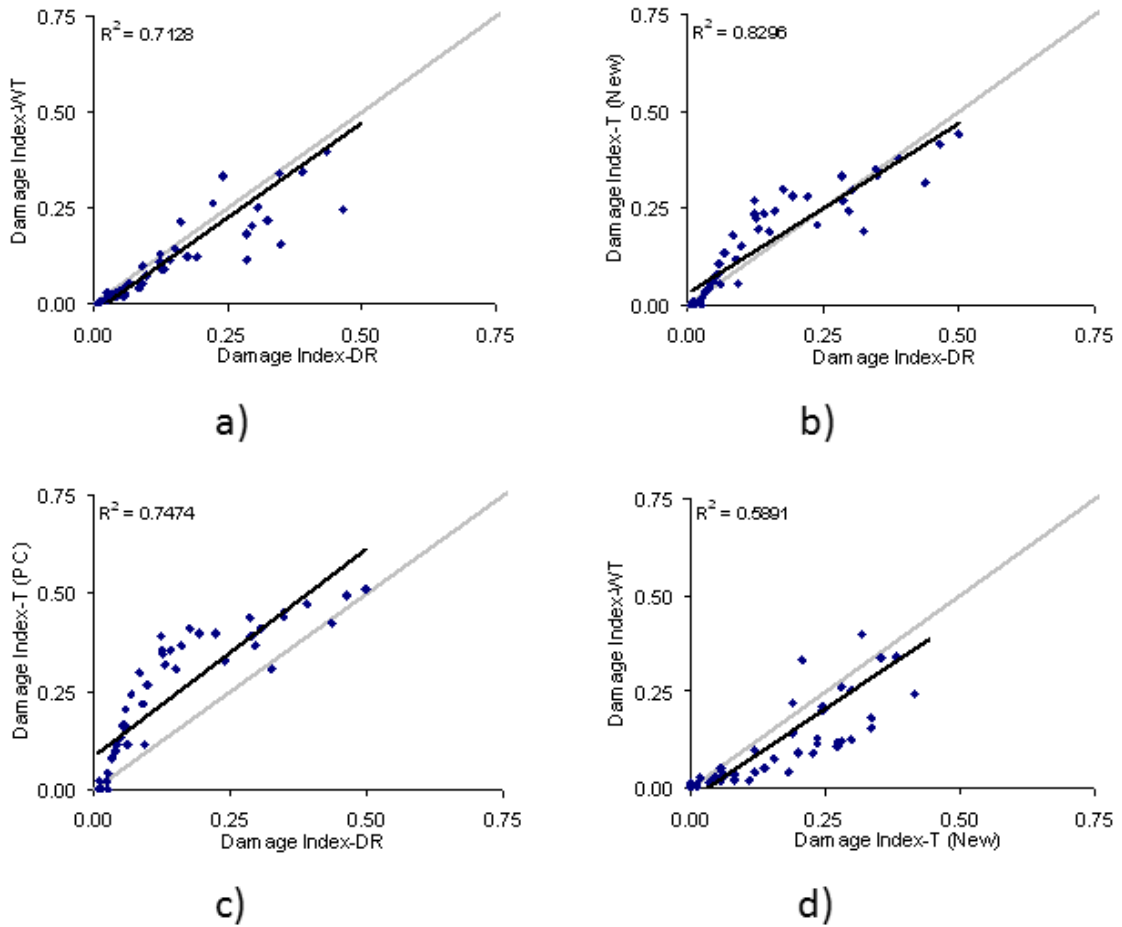

Figure 12. Correlation between different damage indices reference to the third building.

The gray solid line in each figure indicates the linear correlation between the damage indices while the black solid line indicates the linear regression curve fitted to the data points. Figure 10(a), 11(a) and 12(a) show that the inter-story drift based damage index correlates linearly with the wavelet energy based damage index similar to the results in the first section. However, a significant scatter can be observed at higher damage levels for the first building. Some earthquakes induce higher level of wavelet energy based damage while they induce lower inter-story drift based damage and conversely, some earthquakes induce low level of wavelet based damage while they induce higher inter-story drift based damage. This is because the wavelet energy based damage index can account for the effects of number of inelastic cycles and the level of maximum displacement to the damage of a structure while inter-story drift based damage index can only account for the effects of maximum inter-story displacement. However, linear regression lines matched with linear correction lines highlights that the effects of number of inelastic cycles to the damage can be minimized for gravity design buildings considering the average response. Furthermore, it is important to note that the scatter is insignificant at lower inelastic deformation levels highlighting the fact that the effects of number of inelastic cycles to the damage are not significant at lower inelastic deformation levels.

Similar to the wavelet energy based damage index, Figure 10(b), 11(b) and 12(b) also show that a good correlation exists between the inter-story drift based damage index and the inelastic period based damage index proposed in this study for gravity design frames. But,

Figure 10(c), 11(c) and 12(c) indicate significant under estimation of the damage at higher damage levels by the inelastic period based damage index proposed by [10] Di Pasquale and Cakmak (1990) compared to the inter-story drift based damage. This is due to the fact that a structure reaches the damage level 1 at the infinity period of the structure, if the damage level 1 is considered as the ultimate limit state of the structure. However, this predicts the lower and medium levels of damage quite similar the other damage indices for the frame structures.

## 5  CONCLUSIONS

This study proposed new two damage indices based on the wavelet energy and the inelastic period to account the effects of number of inelastic cycles to the damage of a structure. From the damage assessment results of gravity design buildings, linear correlations between the inter-story drift based damage index and the wavelet energy based damage index, and inelastic period are observed. Therefore, it can be concluded that the drift based damage index can be used for the damage assessment of gravity design building in which relatively lower inelastic deformation capacity exist with adequate accuracy ignoring the effects of number of inelastic cycles.

The damage levels for the assessment of a gravity design building using either of wavelet energy or inter-story drift based damage index must be specified according to the failure modes of element which lead to the global failure. However, this study considers only two damage levels: serviceability and ultimate limit state. Therefore, if the element failure in low ductile flexure mode leads to the global failure of a gravity design frame, the wavelet energy or inter-story drift based damage index at the serviceability limit state can be taken to be equal to 0.25 while it is equal to 1 at the ultimate limit state. Furthermore, if the element failure in brittle shear or combination of flexure and shear mode leads to the global failure of a gravity design frame, they can be taken to be equal to 0.5 at the serviceability limit state while they are equal to 1 at the ultimate limit state.

Furthermore, this study highlights the importance of accounting the effects of the number of inelastic cycles to the damage assessment for the structures which have high ductile failure mode as in the case of ductile concrete or steel moment resisting frames. However, further study is required to establish the correlation between the damage parameters. In order to apply these issues in the framework of Energy-Based Design, the range of values of the Energy dissipation Capacity should be put in perspective with respect to the range of values of Input Energy spectra.